\documentclass[a4paper,11pt]{article}
\usepackage{aaskaiid}

\title{Machine Learning and the SKA for Cosmic Dawn and the Epoch of Reionization}
\ShortTitle{SKA Machine Learning}

\author[7,13]{Anshuman Acharya}
\author[9,10]{Michele Bianco}
\author[11]{Daniela Breitman}
\author[18]{Huaxi Chen}
\author[1]{Abhirup Datta}
\author[13,14]{Kangning Diao}
\ShortName{SKA EoR Science Working Group} 
\author[2,3]{Sambit K. Giri}
\author[12]{Caroline S. Heneka}
\author[4]{Nicholas Kern}
\author[19]{Adrian Liu}
\author[1]{Yashrajsinh Mahida}
\author[1]{Suman Majumdar}
\author[1]{Samit Kumar Pal}
\author[18]{Shulei Ni}
\author[12]{Yannic Pietschke}
\author[5]{Davide Piras}
\author[6,7]{Abinash Kumar Shaw} 
\author[15,16,17]{Hayato Shimabukuro}
\author[14]{Ce Sui}
\author[1]{Anshuman Tripathi}
\author[8]{Xiaosheng Zhao}

\affiliation[1]{Department of Astronomy, Astrophysics \& Space Engineering, Indian Institute of Technology Indore, Indore 453552, India}
\affiliation[2]{Department of Astronomy and Oskar Klein Centre, AlbaNova, Stockholm University, SE-10691 Stockholm, Sweden}
\affiliation[3]{Van Swinderen Institute for Particle Physics and Gravity, University of Groningen, Nijenborgh 3, 9747 AG Groningen, The Netherlands}
\affiliation[4]{Department of Physics, University of Michigan, 450 Church St., Ann Arbor, MI, 48109-1040, USA}
\affiliation[5]{Département de Physique Théorique, Université de Genève, 24 quai Ernest Ansermet, 1211 Genève 4, Switzerland}
\affiliation[6]{Department of Computer Science, University of Nevada Las Vegas, 4505 S. Maryland Pkwy., Las Vegas, NV 89154, USA}
\affiliation[7]{Max-Planck-Institut für Astrophysik, Garching D-85748, Germany}
\affiliation[8]{Department of Physics \& Astronomy, The Johns Hopkins University, Baltimore, MD 21218, USA}
\affiliation[9]{Laboratoire d'Astrophysique, École Polytechnique Fédérale de Lausanne (EPFL), Observatoire de Sauverny, Chemin Pegasi 51, CH-1290 Versoix, Switzerland}
\affiliation[10]{Institute for Particle Physics and Astrophysics, ETH Zurich, Wolfgang-Pauli-Str 27, CH-8093 Zurich, Switzerland}
\affiliation[11]{Scuola Normale Superiore (SNS), Piazza dei Cavalieri 7, Pisa, PI, 56125, Italy}
\affiliation[12]{Institut für Theoretische Physik,
Universität Heidelberg, Philosophenweg 16, 69120 Heidelberg, Germany}
\affiliation[13]{Berkeley Center for Cosmological Physics, University of California, Berkeley, CA 94720, United States}
\affiliation[14]{Department of Astronomy, Tsinghua University, Beijing 100084, China}
\affiliation[15]{South-Western Institute for Astronomy Research (SWIFAR), Yunnan University, Kunming, Yunnan 650500, People's Republic of China}
\affiliation[16]{Key Laboratory of Survey Science of Yunnan Province, Yunnan University, Kunming, Yunnan 650500, People's Republic of China}
\affiliation[17]{Nagoya University, Graduate School of Science, Division of Particle and Astrophysical Science, Chikusa-Ku, Nagoya, 464-8602, Japan}
\affiliation[18]{Research Center for Computational Earth and Space Science, Zhejiang Laboratory, Hangzhou 311121, China}
\affiliation[19]{Department of Physics and Trottier Space Institute, McGill University, 3600 University Street, Montreal, QC H3A 2T8, Canada}
\emailAdd{adrian.liu2@mcgill.ca, mbianc@phys.ethz.ch, heneka@thphys.uni-heidelberg.de}

\abstract{When operational, the SKA will generate unprecedented amounts of data and provide exquisite sensitivity for $21\,\textrm{cm}$ tomography of Cosmic Dawn (CD) and the Epoch of Reionization (EoR). With this comes opportunities for new data-driven algorithms that unlock new methods for instrument modelling, data analysis, theoretical simulation, and inference for understanding the high-redshift universe. In this chapter, we provide an overview of some machine learning algorithms that have been proposed for CD and EoR science with the SKA}

\begin{document}
\maketitle

\section{Introduction}
\label{sec:intro}
The Square Kilometre Array (SKA) promises to revolutionize the study of Cosmic Dawn (CD) and the Epoch of Reionization (EoR). A series of impressive precursor experiments (such as PAPER, MWA, LOFAR, HERA) have set an increasingly stringent set of upper limits on $21\,\textrm{cm}$ emission from these epochs, and the SKA provides the necessary sensitivity to not only \emph{detect} but also to \emph{characterize} the cosmological signal to high significance. By simply using conventional data analysis techniques to produce conventional summary statistics such as the power spectrum, the SKA's unprecedented sensitivity is already expected to deliver world-leading constraints on the astrophysics governing CD and EoR.

Going beyond quantitative differences, the SKA pushes $21\,\textrm{cm}$ studies of CD and EoR into qualitatively new regimes. SKA data rates, for example, will be larger than for any radio telescope to date and will necessitate real-time processing that has largely been an optional convenience for current observatories. In terms of science products, the SKA is expected to move beyond summary statistics and to produce images of CD and EoR, enabling a new and incisive set of studies.

With new challenges come new opportunities, and in this chapter we examine the way in which the development of the SKA has been coincident with another fast-developing field---machine learning (ML). Given the relatively recent rise of artificial intelligence techniques in astronomy, it would be fair to say that it is as-yet unclear what the ultimate impact of ML on the SKA will be. However, in this chapter we provide a speculative look at a sampling of ideas that have already circulated in the literature. These ideas span the full breadth of activities of the SKA, from low-level data analysis algorithms to new ways to extract CD and EoR science.

\section{Observations}
\label{sec:obs}
A key obstacle to the detection and characterization of CD and EoR is in understanding the interplay between instruments and sky emission. In other words, what is crucial is the understanding of the cosmological \emph{as seen by the instrument}. This has led to efforts both in simulating instruments to high precision and in the development of novel analysis methods. In this section, we examine the various areas in which machine learning has led to new ways of bridging the instrument-sky divide.

\subsection{Radio Frequency Interference}\label{sec:rfi}

Radio Frequency Interference (RFI) represents a major obstacle to the detection and characterization of the cosmological 21 cm signal and other faint radio sources targeted by the SKA. RFI arises from a wide variety of terrestrial and satellite-based transmitters, and its highly variable nature across both time and frequency domains can mimic or obscure genuine astronomical signals. Effective mitigation of RFI contamination is therefore essential for preserving the scientific integrity of the data and enabling precision measurements of cosmological and astrophysical phenomena.

A number of traditional approaches have been developed to identify and suppress RFI in radio observations. Early techniques relied on the statistical detection of outliers in the time–frequency domain, employing algorithms such as SumThreshold \citep{SumThreshold} and CUMSUM \citep{CUMSUM}. These methods typically detect and flag RFI by identifying significant deviations from local noise statistics and have been implemented in widely used software such as AOFlagger \citep{AOFlagger}. Other approaches have exploited statistical estimators such as spectral kurtosis \citep[e.g.,][]{RFI_kurtosis} to detect non-Gaussian signal components, or applied baseline fitting and harmonic analysis to suppress large-scale or periodic interference \citep[e.g.,][]{RFI_periodic}. While these classical techniques are reliable for specific RFI morphologies, they are often constrained by heuristic thresholds and limited sensitivity to complex or evolving interference patterns.

With the rapid growth of data-driven astronomy, deep learning (DL) models have emerged as powerful alternatives that can operate directly on preprocessed spectrograms or visibility data, reducing the need for manual feature engineering. The U-Net architecture \citep{RFI-Unet}, originally designed for biomedical image segmentation, was among the first deep models applied to RFI detection. Building on this foundation, subsequent studies introduced variants such as RFI-Net \citep{RFI-rfinet}, which incorporated residual connections, and DSC-Dual-ResUNet \citep{RFI-duelresnet}, which employed dual U-Net structures with depthwise separable convolutions to enhance efficiency. \citet{2023MNRAS.520.5552P} used a U-Net architecture to tackle an RFI-related task: after RFI flagging, some analysis treatments \emph{in-paint} the resulting gaps with statistically representative synthetic data. \citet{2023MNRAS.520.5552P} examined the robustness of using U-Nets to performing inpainting, performing a comparison with other techniques.

Other works have explored recurrent neural networks (RNNs) and long short-term memory (LSTM) layers to capture temporal correlations in dynamic RFI signals \citep{RFI-RNN}. Beyond architectural innovations, \citet{RFI-RNet} also investigated alternative training strategies, including transfer learning, to enhance model performance and generalization. Collectively, these efforts demonstrate the ability of DL architectures to learn complex RFI patterns directly from large datasets, achieving superior accuracy and adaptability compared to traditional methods.

Beyond supervised learning, several groups have pursued unsupervised and generative approaches to address the scarcity of labeled RFI data. Generative adversarial networks (GANs) have been trained to model uncontaminated background data and identify anomalies corresponding to RFI events \citep{RFI-modelcomparison}, while convolutional autoencoders (CAEs) have been used to learn latent representations of clean data for anomaly detection \citep{RFI-CAE}. In one example, \citet{RFI-AEKNN} combined CAE-based latent representations with KNN clustering to identify RFI in contaminated observations. Similarly, clustering algorithms such as DBSCAN have been applied to group false-positive events in reduced feature spaces, effectively identifying outlier patterns associated with interference \citep{RFI-DBSCAN}. These approaches highlight the growing shift toward data-driven, model-agnostic strategies capable of adapting to diverse observational conditions without explicit supervision.

Despite these advances, several challenges remain. ML-based models—particularly deep architectures—can be sensitive to the domain and distribution of their training data. Their performance often degrades when applied to datasets with differing instrumental configurations, observing conditions, or spectral resolutions \citep{RFI-rfinet,RFI-modelcomparison}. Furthermore, the deployment of ML methods in large-scale pipelines raises practical concerns about computational efficiency, interpretability, and the need for continual retraining as the RFI environment evolves.

While ML methods are promising, most operational pipelines still rely primarily on classical statistical flaggers, with ML approaches largely remaining at the validation or complementary stage.
Continued progress will depend on developing domain-adaptive, interpretable, and computationally efficient ML architectures that can be seamlessly integrated into the SKA’s real-time data processing framework.

\subsection{Ionospheric Effects} \label{sec:ioneff}

The ionosphere, a turbulent, low-density, multispecies ion plasma that surrounds the Earth, is driven mainly by solar extreme ultraviolet radiation and is a known source of direction-dependent effects (DDEs) in low-frequency radio astronomy. These DDEs introduce systematic errors that fundamentally limit on the dynamic range of radio interferometric observations, particularly for experiments targeting the CD-EoR signal. For a comprehensive discussion of the mitigation of ionospheric errors with SKA-Low, we refer the reader to the \citet{SKAforegrounds}. The dynamic nature of the ionosphere produces flicker noise \citep{2014Datta} ($1/f$ noise, where $f$ is the dynamical frequency). These errors do not reduce with long integration.\footnote{Some reduction will naturally result from stacking multi-night data, which can cause a decorrelation between night-to-night independent components. However, systematic effects (e.g., calibration errors) may still be present.} Hence, there is a need for ionospheric calibration at shorter time intervals that match the ionospheric variability. The residual DDEs or the residual ionospheric refractive shifts have, in some studies, been a non-negligible component in residual excesses. the upper limits of the CD/EoR power spectrum measurements \citep{Pal2025}. Therefore, mitigating these residual DDEs is essential for robust interpretation of the faint CD-EoR signal in both the Fourier and image domains. For example, a recent study by \citet{Brackenhoff2024} demonstrated that the excess noise induced by ionospheric phase errors can be removed using Gaussian Process Regression (GPR; see Sec.~\ref{sec:foregroundmitigation} for more details and another application). They found that the propagation of ionospheric errors from longer to shorter baselines, which are the most affected by induced phase errors, may not be a dominant contributor to the excess noise in this LOFAR study.

In \citet{Tripathi2024}, an artificial neural network (ANN)–based framework was developed to recover the global 21-cm signal in the presence of foreground and ionospheric effects. The study found that, for slowly varying ionospheric conditions, a simple ANN architecture is capable of successfully recovering the underlying cosmological signal.\footnote{In essence, the ANN is taking the place of the usual Bayesian inference exercise of fitting a global signal or a power spectrum to infer astrophysical parameters. While currently it is still generally practical to perform Bayesian inference once a summary statistic has been produced, future analyses that incorporate yet more nuisance parameters may benefit from the faster runtime of a machine learning-based algorithm.} Several machine learning frameworks have also been developed to reconstruct the Total Electron Content (TEC) in radio interferometry calibration, correcting the DDEs arising from ionospheric effects \citep{Li2025, Jong2025}. \citet{Albert2020} propose a probabilistic physics-informed model using Gaussian processes to infer ionospheric phase screens, thereby reducing residual errors.

\subsection{Gain Calibration}\label{sec:gain}

The gain calibration errors are expected to be present in any radio interferometric observations of the CD-EoR. Generally, the calibration is done by estimating the Jones matrices with the assumption of a perfectly known model sky, or (in cases where a radio interferometer has many repeated copies of the same baselines) by exploiting the self-consistency of a redundant array. However, limitations such as parameter degeneracies, incompleteness of sky models, imperfect knowledge about the primary beam, and rapid atmospheric fluctuations result in residual calibration errors. These residual errors can lead to incorrect interpretations of the observed signal \citep{Datta2009, Beardsley2016}, introducing spurious features in the image domain \citep{Pal2025b} and biasing statistical estimates of the signal in the Fourier domain, such as the power spectrum \citep{Barry_2016, Kern2020, Byrne_2021, Mazumder2022}. Such biases can propagate through the inference pipeline to the estimation of astrophysical parameters, and thus ultimately lead to misinterpretation of the underlying physical processes \citep{2025Tripathi}. For estimators of signal power spectrum, these calibration errors can introduce excess power across different \(k\)-modes.

Mitigating these errors is therefore crucial for a robust interpretation of the signal in both image and Fourier domains. Motivated by this, various methodologies have been developed for this purpose. For instance, to address the excess power introduced in the power spectrum by residual gain errors, techniques such as Gaussian Process Regression (GPR) and Principal Component Analysis (PCA) along with other unsupervised machine learning methods have been applied to identify the \(k\)-modes within the EoR window most affected by calibration systematics, enabling the construction of more reliable estimators for the signal power spectrum \citep{Mertens2020, Chen2023, Brackenhoff2024, beohar2025}. A recent study by \citet{beohar2025} demonstrates that a hybrid strategy combining avoidance with PCA and GPR enables robust signal recovery across the entire \(k\)-range. They found that GPR and PCA-based gain error correction performs more effectively at large scales, although both methods tend to underestimate the power on smaller scales.

In addition, deep learning-based models, particularly U-Net architectures, have been employed to correct for gain errors arising from frequency-dependent Gaussian beams and bandpass fluctuations, thereby improving 21-cm signal recovery \citep{Chen2024}. \citet{Leeney:2025} proposed a machine learning–based calibration framework for a 21-cm sky-averaged cosmology experiment, aimed to accurately model the complex gain behaviour of the radio instrument.

\subsection{Imaging}\label{sec:imaging}

One of the significant innovations of the SKA-Low telescope is that it will be sensitive enough not only to detect the 21-cm signal but also to produce 3D tomographic maps of its distribution at different stages of the EoR \citep{Wyithe2015ImagingSKA, mellema2015hi, giri2018optimal}. For a comprehensive discussion of 21-cm signal imaging with SKA-Low, we refer the reader to \citet{SKAimaging}. Over the years, several AI methods have been developed to process images and 3D data \citep[e.g.][]{LeCun1997, Baldi2012autoencoders, Ronneberger2015unet, Kingma2022varautoencoding}, so naturally, machine learning techniques have been establishing their role as the primary method for image analysis among cosmologists and astrophysicists \citep[see][for a review]{Dvorkin2022machinelearningcosmology, Lahav2023deepmachinelearningcosmology}. The application of these techniques to 21-cm cosmology has accelerated in recent years, driven by the need to extract maximum scientific information from the complex, high-dimensional tomographic datasets that SKA-Low will produce. In this section, we focus on the imaging process itself, reserving \emph{applications} of imaging to later sections.

A fundamental challenge is addressing the systematic errors in radio interferometric imaging that arise from incomplete $uv$-coverage and, relatedly, the effect of the synthesised beam. These introduce artefacts in radio images that can be helpful to remove (``deconvolution") for many applications, such as mitigating foreground contamination from extragalactic compact or point-like radio sources. Traditionally, CLEAN-based methods have been used for radio interferometric imaging. Initially proposed by \citet{hogbom1974aperture}, \texttt{CLEAN} iteratively deconvolves the telescope beam by identifying and subtracting point sources from the dirty map, with the accumulated model convolved with a restoring beam to form the final image. Over the years, extensions of this method have been suggested, such as the multi-scale, multi-frequency model, implemented in \texttt{MS-CLEAN} and \texttt{MF-CLEAN}, for wide-band image convolution \citep{rau2011multi}, and wide-field techniques such as $w$-projection and faceting, implemented in WSClean \citep{offringa2014wsclean}, correct for direction-dependent effects. Although effective for compact sources, these methods struggle with diffuse or complex foregrounds, as representing sky brightness with discrete components can introduce additional artefacts. Their iterative nature also makes them computationally demanding, especially for wide-field, low-frequency surveys.
An unsuccessful deconvolution process (or at least one whose statistical properties are not well-understood) may exacerbate foreground contamination and increase the difficulty of signal separation~\citep{Hothi:2020dgq, Ni:2022kxn, Wang:2024ept, Cox:2023neq}. This, therefore, motivates the exploration of alternative deconvolution approaches.

\begin{figure}
    \centering
    \includegraphics[width=\columnwidth]{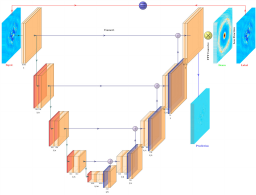}
    \caption{The \texttt{PI-AstroDeconv} architecture. The yellow blocks represent convolutional layers, red blocks represent pooling layers, and gray-green blocks represent upsampling operations. The left half of the network depicts the downsampling path, while the right half represents the upsampling path. The number of channels is indicated at the bottom of each block. The arrows above the image symbolize the connections between network layers. The numbers below each layer in the network indicate the quantity of convolutional kernels present in that specific layer. Additionally, the letters denote the variations in the image dimensions within the layer. The letter "I" specifically represents the size of the input image. The blocks located at the far left and far right of the network indicate the input and the label, respectively (represented as deep blue circles). A Depthwise convolutional layer, represented as a yellow circle, is inserted between the U-Net network and the label. This layer utilizes convolutional kernels resembling telescope beams and is accelerated using FFT. The U-Net's output is convolved with the beam, and the result is compared with the labels (Ground Truth) to compute the loss function. Network inputs and ground truth labels are sourced from the telescope. The model is trained to embed physical priors through convolutional layers, ensuring outputs align with the original observations. The final prediction of the network is derived from the last layer of the U-Net, represented as the last yellow layer. The symbols $+$, $\times$, and $\equiv$ represent concatenation, convolution, and exact equality, respectively.
    }
    \label{fig:deconv}
\end{figure}

To overcome some of the limitations mentioned above, researchers have proposed various AI-based deconvolution approaches. For instance, \cite{Schmidt2022deepimg} and \cite{Geyer2023} employ a convolutional neural network to reconstruct images of compact radio sources by directly inferring missing visibility data from incomplete Fourier-space data, offering improvements in speed, automation, and reproducibility over traditional methods such as \texttt{CLEAN}. Another noteworthy example is \texttt{PI-AstroDeconv}~\citep{ni2024model,ni2025application}, whose network architecture is illustrated in Figure~\ref{fig:deconv}. Rather than relying solely on deep networks for blind learning, PI-AstroDeconv constructs an explicit imaging model based on the physical principles of observational imaging:
\begin{equation}\label{eq:deconv}
    I_D(x, y) = P_D(x, y)\otimes I'(x, y),
\end{equation}
where the dirty map $I_D(x, y)$ is the convolution of the clean map $I'(x, y)$ with the telescope beam function $P_D(x, y)$, thereby explicitly modelling the effect of the beam on the observed image, as illustrated in Figure (\ref{fig:deconv}). Based on this model, \texttt{PI-AstroDeconv} transforms the beam deconvolution problem into an image-to-image mapping task: given a dirty map and its corresponding PSF, the network directly outputs an estimated clean map, which is then convolved with the same PSF to produce a “dirty map” for loss computation. The training minimises only the Log-Cosh distance between simulated and true dirty maps, achieving fully unsupervised learning without requiring clean images as labels.

PI-AstroDeconv operates directly on image-domain data; its inputs are the dirty maps produced by interferometric imaging and the corresponding PSFs, without relying on conventional deconvolution operations. This design substantially simplifies the preprocessing workflow and introduces minimal additional computational overhead. To accommodate the large size and high resolution of radio images, the network performs the final convolution in the frequency domain using FFTs, reducing computational complexity from $\mathcal{O}(n^4)$ to $\mathcal{O}(n^2 \log n)$, significantly lowering memory usage and training time. The convolution kernel is fixed as the input PSF and does not update with gradients, embedding the telescope beam response as a hard constraint in the loss function to ensure physical consistency.

\subsection{Foreground Mitigation}\label{sec:foregroundmitigation}

A key challenge in 21-cm imaging is that astrophysical foregrounds (e.g., synchrotron and free-free emissions from the Milky Way and other galaxies) dominate over the cosmological signal in a wedge-shaped region of the cylindrical Fourier space ($k_\perp$, $k_\parallel$). One way to mitigate foreground is to filter out the wedge region (``foreground avoidance'') from the visibility data before image deconvolution. With this technique, we can focus on areas in the Fourier space that are mostly foreground-free \citep[see][for more details]{Kerrigan_2018, Morales_2019, HERA_2023}. However, this process removes a relevant portion of cosmological information along with the undesired foregrounds. Thus, a worthwhile goal is to reconstruct the 21-cm signal in the wedged region and/or successfully model and remove foregrounds, thereby maximising the physical scales that can be explored with the 21-cm power spectrum and boosting sensitivity by orders of magnitude \citep{Pober_2014}. 

Machine learning methods have been developed to recover the spatial information lost during wedge filtering by exploiting the non-Gaussian correlations in the 21-cm signal \citep{GagnonHartman2021, Kennedy2024mlrec}. Deep generative models based on stochastic interpolants have also been introduced to reconstruct 21-cm data lost to wedge filtering, leveraging the non-Gaussian nature of the signal to map wedge-filtered 3D lightcones to samples from the conditional distribution of wedge-recovered lightcones \citep[e.g.][]{Sabti2024generativemodel, chen2025fieldlevelreco21cm}. Additionally, approaches have been developed to combine standards foreground mitigation strategies with novel AI techniques to recover complete 21-cm signal images directly from foreground-contaminated data. For instance, the work of \cite{Bianco2024deep} combines U-Nets with existing foreground removal methods (e.g., PCA, wedge filtering, polynomial decomposition, and GPR), improving the recovery of both neutral (HI) and ionised (HII) regions from tomographic data. Furthermore, the recovery of the 21-cm signal in tomographic maps has been enhanced by modifying the U-Net architecture to accept an additional input, which serves as prior information. With the addition of a \textit{convolutional intercepting layer}—a series of pooling and convolutional operations that intercept and combine with the output of the skip connections that process images with residual foreground—additional maps are processed to integrate prior information on the neutral/ionised in observed foreground contaminated maps \citep{Bianco2025deep}, demonstrating an enhancement in the recovery of 21-cm tomographic data at small scales and across a wide range of astrophysical parameters and redshifts. These maps can be derived from observations of high-redshift galaxies.

Another approach to machine learning-inspired foreground mitigation has been to use Gaussian Process Regression (GPR). GPR can be employed to model and subtract foreground contamination \citep{Mertens_2018, Gehlot_2019, Hothi_2021}, with each data component being modelled using analytic covariance kernels. However, the use of simplified analytic kernels for the EoR signal can limit the accuracy of component separation. It may bias the recovered 21-cm signal \citep[for example, see][]{Kern_2021}, motivating the development of more flexible, data-driven physics-motivated kernel models. 

Machine learning offers a powerful alternative by learning covariance priors directly from physics-based simulations. In particular, ML-enhanced GPR (ML-GPR) has been recently developed to enhance foreground mitigation \citep{Mertens2024, Acharya2024}. This approach utilises unsupervised generative models (e.g., Variational Autoencoders) trained on simulated 21-cm lightcones from codes such as \texttt{21cmFAST} \citep{Mesinger_2007, Mesinger_2011} or \texttt{GRIZZLY} \citep{Ghara_2015, Ghara_2018} to build flexible, data-driven covariance kernels. These learned quantities then replace the fixed analytic kernels in the usual GPR pipeline. By fitting the latent parameters of the ML-derived kernel to the data, the method mitigates mismatches between assumed and true signal covariance. It can adapt to complex spectral features and non-Gaussianity of the 21-cm signal that rigid kernels miss. According to \citet{Acharya2024}, ML-GPR outperforms standard GPR in component separation and remains robust even when the true signal lies outside the training set, thereby providing resilience to instrumental systematics and noise. This has now been incorporated in foreground mitigation pipelines for various observations, such as with LOFAR (applied to the NCP field in \citealt{Acharya2024b, Mertens2025}, the 3C196 field in \citealt{Ceccotti2025}) and NenuFAR \citep{Munshi2024, Munshi_2025}. Additionally, it has demonstrated the potential to detect the 21-cm power spectrum by isolating the cosmological component not only from foregrounds but also from more complex systematics. In this regard, upper limits on the 21-cm power spectrum, after separating other contributions, have been reported in \citet{Acharya2024b} and \citet{Mertens2025}. 

A key innovation in ML-GPR is the use of a Variational Autoencoder \citep[VAE;][]{Kingma_2013} to learn a latent representation of the 21-cm signal’s covariance structure. A VAE is a generative probabilistic model where observed data $x$ (here, realizations of the 21-cm signal from simulations) are compressed into latent variables $z$ drawn from a prior distribution $p(z)$, typically constrained to be as close as possible to a multivariate Gaussian $\mathcal{N}(0, I)$. The model consists of two neural-network components: an \emph{encoder}, $q_\phi(z|x)$ with parameters $\phi$, which approximates the intractable posterior distribution over the latent variables, and a \emph{decoder}, $p_\theta(x|z)$ with parameters $\theta$, which reconstructs data from latent samples. Training is achieved by maximizing the evidence lower bound (ELBO):
\begin{equation}\label{eq:elbo}
\mathcal{L}(\theta, \phi; x) =
\mathbb{E}_{q_\phi(z|x)} \left[ \log p_\theta(x|z) \right] 
- D_{\mathrm{KL}} \!\left( q_\phi(z|x) \,\|\, p(z) \right).
\end{equation}
In Eq.~\ref{eq:elbo}, the first term is the expected log-likelihood of reconstructing the data under the generative model, encouraging accurate recovery of $x$ from its latent representation. The second term is the Kullback--Leibler divergence, which penalizes deviations of the approximate posterior $q_\phi(z|x)$ from the prior $p(z)$. This regularization enforces smoothness and continuity in the latent space, preventing overfitting and enabling meaningful interpolation between latent variables.


In the ML-GPR framework \citep{Mertens2024, Acharya2024}, only the encoder is retained after training. The encoder provides a mapping from high-dimensional 21-cm signal realizations into a structured latent space that captures their statistical correlations. From this representation, one can construct a data-driven covariance kernel, \begin{equation}\label{eq:covarkern}
k_{\mathrm{ML}}(x, x') \approx
\mathbb{E}_{z \sim q_\phi(z|x, x')} \Big[ (x - \mu_z)(x' - \mu_z)^\top \Big],
\end{equation} where $\mu_z$ denotes the mean of the encoded latent distribution. This learned kernel $k_{\mathrm{ML}}$ replaces hand-crafted analytic kernels in GPR. In practice, the VAE encoder learns the manifold of possible EoR signals from simulations, such that the covariance structure naturally reflects both astrophysical variability (e.g., ionization histories, source models) and instrumental effects (e.g., chromaticity, mode-mixing). This enables ML-GPR to remain flexible to realistic deviations, in contrast to standard GPR approaches where the covariance must be specified a priori.


In summary, ML-GPR leverages modern generative models to extend classical GPR for 21-cm analysis. By learning covariance kernels from simulations reduces prior mismatch and signal loss, adapts to non-ideal spectral features, and yields more accurate 21-cm reconstructions. These advances have already improved foreground pipelines in deep EoR fields, and further improvements (learnt kernels for the foregrounds, other ML algorithms, etc. apart from improvements in data reduction) are also in development. All these improvements promise a robust path toward detecting the cosmic 21-cm signal with the SKA-Low. Indeed, in the recent SKA Science Data Challenge 3a, foreground (SDC3a\footnote{\url{https://sdc3.skao.int/challenges/foregrounds}}), participants were asked to recover the 21-cm cylindrical power spectra from foreground contamination and systematics from visibility and/or tomographic data, and GPR resulted in the smallest residuals and least bias \citep{Bonaldi2025}.
\section{Theory}
\label{sec:theory}
In addition to experimental concerns, machine learning methods have also had an impact on theoretical calculations. In many ways, CD/EoR represent interesting regimes to apply machine learning, since (unlike many other sub-fields in cosmology) there are both computational limitations and considerable uncertainties as to the correctness of models.

\subsection{Accelerating \& Improving Simulations at the Field Level}\label{sec:simulations}

Numerical simulations are a key tool for studying the Epoch of Reionisation (EoR). Yet, reionisation is an intrinsically multi-scale problem: simulations must encompass cosmological volumes of order $\sim 500\,\mathrm{Mpc}$ to capture the large-scale topology of ionised regions \citep{Iliev2014howlarge}, while simultaneously resolving sub-galactic processes that shape the ionising sources \citep{Trebitsch2017fluct}, over an extended period of cosmic history from $z \sim 20$ to $5$ \citep{GiriBianco2024}. Meeting these requirements makes such simulations computationally expensive, demanding billions of CPU and/or GPU hours on thousands of nodes in the world’s most powerful supercomputers \citep[e.g.][]{Ocvirk2016codaI, Ocvirk2020codaII, Trebitsch2021obelisk, Garaldi2022}. To mitigate these limitations, several EoR simulations have been adapted to run on GPU devices, enabling significant performance improvements and speed-ups \citep{Aubert2010aton, Aubert2015emma, Hirling2024pyc2ray}. However, despite these advances, such simulations remain computationally demanding. For this reason, recent studies have begun exploring AI techniques to overcome the computational constraints inherent in EoR simulations.

When integrating AI with EoR simulations, two main objectives can be identified. The first focuses on replacing the radiative transfer solver with a learned model that emulates its output, predicting quantities such as the hydrogen neutral fraction or ionisation field from input data describing the ionising sources (e.g. positions, intrinsic luminosities, and UV mean-free paths), the IGM density distribution, or directly from a set of astrophysical parameters. The second aims to accelerate specific components of the simulation pipeline by employing neural networks that learn the complex relationships between astrophysical processes (e.g. star formation history, stellar radiative feedback) and their host environments (i.e. dark matter haloes and/or overdensities in the IGM) on scales that are generally not resolved in reionisation simulations.

An example of the first application was presented by \cite{Chardin2019deep}, who employed a convolutional autoencoder that takes 2D images of the gas density field and the stellar number density from a simulated lightcone and reconstructs the reionisation time, $t_\mathrm{reion}(\mathbf{r})$—the cosmological time at which a region reaches a given ionisation fraction threshold. Two separate encoders process the two inputs, and an averaging operation combines the two low-dimensional latent spaces before deconvolutional operations are applied. This method shows promising results for large-scale features ($\geq 10,\mathrm{cMpc}$) but struggles to recover small-scale structures and tends to delay the appearance of the first ionised regions at high redshift. In contrast, \cite{Korber2023PINION} proposed a different approach that focuses on recovering small-scale details of reionisation. In this work, a CNN processes input data from sub-volumes of a larger EoR simulation and predicts the neutral fraction of the central voxel. The training is enforced voxel-by-voxel using a physically informed loss function that approximates the hydrogen chemistry equation. The method employs the same assumptions and input fields, but introduces an additional preprocessing step that generates a smoothed source field based on an empirical redshift-dependent model of the mean free path of ionising photons, $\lambda_\mathrm{mfp}(z)\propto\frac{c}{H(z)}(1+z)^{-2.55}$. The resulting field is then used as an extra input to the network. This approach achieves high accuracy on small scales but shows reduced large-scale performance at high redshift due to the mean-free-path assumption. Neither of the methods described above imposes a specific EoR model during training. Their central assumption is that a neural network can learn the underlying physical relationship between the sources and sinks of radiation and the resulting ionised field. In this sense, the network maps the redshift evolution of the ionisation front from the given source distribution, providing an implicit, data-driven approximation of radiative transfer based on pre-run numerical simulations.

Following a similar motivation, \citet{Posture_2025} has integrated U-Net with a Vision Transformer-based architecture (\texttt{CosmoUiT}) to emulate the 3D ionisation field and the corresponding 21-cm field from the EoR. It takes 3D dark matter and halo fields, along with reionisation model parameters, as inputs. In this architecture, the multi-head self-attention mechanism of the transformer captures long-range dependencies, while the convolutional layers in the U-Net capture small-scale variations in the field. Furthermore, the reionisation parameters are provided during training to both the vision transformer block and the U-Net block, conditioning the output on these parameters. The hybrid design of the emulator overcomes the limitations of earlier approaches, capturing both large- and small-scale features of the 21-cm signal, and limiting the error at the boundary between ionised and neutral regions that leads to over-estimation of bubble size distribution at the early stage of reionisation.


Another approach consists of generating 21-cm tomographic maps directly from a set of astrophysical and cosmological parameters. These methods typically assume a specific model that parametrises the sources of radiation and recombination in the IGM, and are generally trained on large datasets ($\sim 10^4$ realisations) to cover the broad parameter space. As such, they can be regarded as emulators trained on a particular model of reionisation and are often integrated into pipelines for EoR parameter inference. Examples include \cite{Chardin2019deep, Korber2023PINION,Zhao:2023, Hothi25, Posture_2025,Heneka:2025}. \cite{Zhao:2023} employs two kinds of generative models to emulate 21-cm image slices: Generative Adversarial Networks (GANs, \citealp{Goodfellow2014}) and diffusion models \citep{SohlDickstein2015, Ho2020}. While GANs are more computationally efficient, diffusion models have been shown to produce a greater diversity of samples, as quantified by scattering transform coefficients. The main limitation of this approach is that 21-cm slices in the training set are limited to high redshifts, where the 21-cm signal exhibits little evolution and remains close to a Gaussian density field. As a result, the generative models often struggle to learn meaningful parameter dependencies from such data. The work of \cite{Heneka:2025} employs large language models (LLMs) to analyse and generate 21-cm tomographic maps from astrophysical and cosmological parameters. In this study, pre-trained transformers (e.g. \texttt{Qwen2.5}) are adapted for out-of-domain physical inference tasks, highlighting an intriguing analogy between natural language and cosmological structure formation. The model accepts a sequence of $\delta T_b$ maps conditioned by redshift, where each slice is flattened into a sequence of patches separated by newline tokens. Similar to a text-generation task, each pixel is treated as a token, and the model learns to predict the next token both spatially and across redshifts. However, such LLM-based approaches struggle to generalise beyond the dynamic range of their training data and often fail to generate coherent 3D 21-cm maps. A general limitation of these applications is that most map emulators are trained on computationally inexpensive simulations, such as small-scale \texttt{21cmFAST} runs. Therefore, training these generative models from scratch typically requires $10^5$–$10^6$ samples, which is prohibitive for high-fidelity simulations. To address this, \citet{Diao_2025} proposed a pre-training and fine-tuning scheme in which a GAN pre-trained on small-scale simulations is fine-tuned to generate large-scale 21-cm maps using only $\sim 10^2$ high-cost simulations.

The second class of AI applications to EoR simulations aims to accelerate or augment the modelling of physical processes that are too time-consuming or computationally expensive to resolve directly. These models are typically trained on high-fidelity simulations or constrained by galaxy observations, enabling them to capture sub-galactic-scale processes that would otherwise require prohibitively high resolution. One of the first examples was presented by \cite{Sullivan2017}, who employed an artificial neural network to model radiative feedback in galaxies due to the UV background radiation. The network successfully connected the evolution of the baryon fraction and the suppression of star formation in galaxies during reionisation to local galaxy properties, including tidal forces, hydrogen ionisation fraction, gas temperature, and the halo virial ratio. Another example is provided by \cite{Sherwin2023}, who used a fully connected neural network to link dark matter halo merger histories and galaxy properties with the UV escape fraction derived from high-redshift galaxy formation simulations. Deep neural networks have also been proposed to predict the stochastic star formation history of galaxies from dark matter halo properties at lower redshifts ($z \sim 4$) \citep[e.g.][]{Moster2021galaxynet, Behera2025predgalform}, and such approaches could be extended to higher redshifts. For instance, \cite{Feathers2025} employed a neural network as a sub-grid model to self-consistently determine the formation of the first stars in minihaloes ($M_\mathrm{halo}\leq 10^8\,M_\odot$) in the presence of a Lyman–Werner background during the early stages of reionisation. The network input consists of parameters from the surrounding large-scale environment of minihalos—such as the IGM overdensity and velocity field—and predicts the star formation rate in each simulation cell, significantly improving accuracy compared to semi-analytical models.

Moreover, the need for large cosmological volumes to reduce sample variance often limits the spatial resolution of simulations. Moreover, the minimum halo mass resolved is typically well above the expected threshold for haloes hosting star-forming galaxies. Here, AI-based super-resolution techniques play an increasingly important role, as they can be employed as alternatives to sub-grid models by artificially increasing the effective spatial or mass resolution of large-volume EoR simulations. The work of \cite{Schaurecker2022superresdm} employs a GAN architecture to populate low mass-resolution N-body simulations with additional smaller haloes, effectively lowering the minimum resolved halo mass by up to an order of magnitude in $\mathrm{cGpc}$ volumes N-body simulations. The resulting halo number counts agree well with theoretical predictions for a given cosmology. However, due to memory constraints, the network input consists of haloes being smoothed on increasingly fine grids rather than producing extended halo catalogues, meaning that the inpainted halo population remains distributed on a per-pixel level—thus limiting the achievable mass-resolution enhancement. Conversely, \cite{Pochinda2025supres} demonstrated that diffusion-based models can improve the resolution of cheap, low spatial-resolution simulations. Their score-based generative framework, combined with stochastic differential equations, can refine coarse but large volume 21-cm simulations to simultaneously reproduce the expected field of view and angular resolution of SKA-Low observations. While the approach recovers the 21-cm power spectrum with marginal errors below the anticipated SKA-Low noise level, it struggles to reproduce the sharp $\delta T_b=0,\mathrm{mK}$ peak associated with ionised regions in the bi-modal 21-cm brightness distribution.

Overall, the integration of AI techniques into EoR simulations offers a promising pathway to bridge the gap between accuracy and computational feasibility. Neural networks provide a fast and cost-effective alternative for estimating global quantities, such as the volume-averaged neutral fraction or the timescale of reionisation, that are typically derived from high-fidelity but computationally expensive simulations. Although data-driven emulators and neural sub-grid models cannot yet fully replace radiative transfer and hydro-dynamical simulations, they have the potential to substantially accelerate parameter-space exploration and aid the calibration of numerical simulations. In the near future, combining carefully trained neural networks with traditional simulations may enable optimised inference from observations, ultimately enhancing our ability to interpret forthcoming 21-cm data from next-generation instruments such as SKA-Low.


\subsection{Emulators of Summary Statistics}\label{sec:emulators}

Observations can be robustly translated into constraints on astrophysics and cosmology via Bayesian inference (for more details see the Bayesian inference section in the Inference chapter; \citealt{Acharya01.2026.SKA}). State-of-the-art Bayesian inference requires multiple thousands of forward model evaluations (e.g., $\sim$200k in \citealt{HERA_2023}). Even the fastest semi-analytical models of the EoR (e.g., \citealt{Santos10, Hutter21,Davies25}) take $\sim$ 1h per evaluation, which can make such inferences computationally impractical. An emerging alternative to direct simulation has been the \textit{emulation} of summary statistics. In contrast to the previous section where one was emulating \emph{fields} (e.g., the $21\,\textrm{cm}$ brightness temperature field), in this section we discuss the use of surrogate models, often in the form of machine learning architectures like neural networks (NNs) to produce outputs such as power spectra from a simulator given a set of input parameters $\theta$. Emulators were first adopted within cosmology (e.g. see \citealt{Heitmann2006, Heitmann2009, Fendt2007, Auld2007}) and have since been widely adopted in astrophysics (e.g. see \citealt{Kern_2017, Shimabukuro_2017}). Emulation requires simulations but allows for \textit{amortized} (i.e., reusable) inference. We still need to pay an upfront computational cost when we generate a suite of simulations to train the emulator (although online emulators have been developed in cosmology as well, see e.g. \citealt{Gunther2025}). This set of simulations, however, is less computationally expensive than performing full inference with an expensive simulator, and provides full flexibility when running several inference rounds is required. The trained emulator can then be reused to quickly interpret observations as soon as they are released and test various theoretical models. Note that there is no \emph{a priori} method to determine the number of training samples required to train a NN to a desired accuracy and precision. Recent work \citep{2024arXiv240204355L} proposed a likelihood-free statistical test, PQMass, which quantifies whether two sets of samples---e.g., emulator outputs and simulations---are drawn from the same distribution, offering a principled way to assess generative model fidelity and sample sufficiency, and can therefore serve as a validation test of whether the training sample size is sufficient. As such, as instruments improve, more accurate and precise emulators will be required, which in turn shall become more and more expensive to train. To tackle this problem, the use of pre-trained models or foundational models which are trained on millions or even billions of observations is gaining popularity \citep{Parker2025}, as reusing the information from previous runs can significantly lower the required computational budget.

Many current 21-cm instruments are focused on a detection of the spherically-averaged 1D 21-cm PS as it is a binned statistic that has enhanced signal-to-noise ratio in comparison to 3D maps. As such, the first emulators in the field naturally targeted the 1D PS (e.g. see \citealt{Kern_2017, Shimabukuro_2017, Schmit18, Jennings_2019, Ghara20, Mondal21, Choudhury2025}).
These models typically emulated the 1D PS to about 10\% accuracy. After a claimed detection from EDGES of the globally-averaged 21-cm signal \citep{Bowman18}, there has been a demand to emulate the global signal as well (e.g. see \citealt{Cohen20,Bevins21, Bye22}). These emulators have sub-percent accuracy despite having generally increased the dimensionality of the parameter space compared to previous efforts. In some cases emulator efforts have incorporated more sophisticated NN architectures such as recursive neural networks (RNNs) \citep{Prelogovic23} and long-short-term memory (LSTM) \citep{DorigoJones24}. With the ever-increasing sensitivity of the SKAO, the number of parameters to constrain will further increase, calling for novel techniques such as those relying on differentiable pipelines, which typically allow for a more efficient exploration of high-dimensional spaces.

The 21-cm signal is inherently non-Gaussian, as its evolution is governed by a complex interplay of highly non-linear astrophysical processes. Unlike the CMB, whose fluctuations are fully described by the power spectrum, the 21-cm signal contains rich higher-order correlations that encode additional information about the underlying physics. Consequently, the power spectrum---while a convenient and widely used summary statistic---represents an imperfect compression of the full 21-cm field and inevitably discards valuable information. To address this limitation, recent studies have developed emulators targeting higher-order statistics such as the bispectrum, which are more sensitive to the non-linear structure of the signal. 

Bispectrum computations are computationally challenging, especially when one is trying to compute all different shapes of unique triangles \citep{Bharadwaj2020, Shaw2021}, and when considering the large variation in dynamic range. A bispectrum emulator is therefore an attractive option for generating a plethora of bispectra samples in future Bayesian analysis of the signal. Recently, \citet{Tiwari_2022} emulated the EoR 21cm bispectra using NNs, and used their emulator to forecast the astrophysical constraining power of bispectra.
    


One drawback to many emulators in the literature is that they produce only point value predictions of the target signal statistic; thus, they fail to capture the uncertainty in their predictions. Therefore, when such emulators are used in the Bayesian inference pipeline, they cannot naturally propagate their prediction uncertainties to the estimated model parameters. One way to propagate these errors is by using GPR to interpolate over errors that are cross-validated over a test dataset \citep{Kern_2017}. The resulting emulator error covariances are then incorporated into one's downstream analyses such as astrophysical parameter inference.
A more sophisticated way is to develop an emulator using Bayesian neural networks (BNN), which can provide the uncertainty associated with its prediction as a model output. In \citet{Mahida_2025}, power spectrum and bispectrum emulators were developed using BNN architectures. These emulators provide the Gaussian posterior prediction of the predicted signal statistic in the form of the mean and covariance as model output. Thus, one can propagate this model prediction uncertainty to the Bayesian inference pipeline very easily.

\section{Inference}
\label{sec:inference}
Having discussed ML improvements to instrument modelling and data analysis in Sec.~\ref{sec:obs} and improvements to theoretical efforts in Sec.~\ref{sec:theory}, we now bring the two together by considering the final step in any observational effort---the \emph{inference} of how a particular astrophysical system (in our case, CD and EoR) behaves.

\subsection{Auxiliary Data Products from Images}\label{sec:aux}

One application of ML models is to perform image segmentation to extract specific astrophysical features from tomographic data. This approach treats the map as an image to be classified pixel-by-pixel, enabling direct characterisation of the morphology and topology of structures during reionisation. U-shaped neural network architectures have been trained to produce binary maps that identify and delineate the morphology of HI and HII regions during the EoR \citep{Bianco2021deep, Bianco2024deep}, even after brightness temperature maps have been distorted by foreground filtering \citep{GagnonHartman2021,Kennedy2024mlrec}. This allows for a direct study of the topology of reionization---such as the size distribution, shape, and connectivity of ionized bubbles---a key science goal for the SKA-Low that can provide insights into the nature and distribution of the first ionizing sources \citep[e.g.,][]{giri2018bubble,giri2019neutral}. Importantly, these segmentation methods present a way to estimate the ionization history in a model-independent manner, as the volume of ionized or neutral regions can be directly measured from the identified features in the images, providing constraints on the reionization timeline without requiring assumptions about the underlying astrophysical processes.

Auxiliary data products arising from ML-processed images can also facilitate synergies with other probes of CD/EoR. For example, \citet{Kennedy2024mlrec} built on the work of \citet{GagnonHartman2021} in recovering ionized bubbles in the presence of wedge-filtered $21\,{\rm cm}$ maps, generalizing the technique to operate on light cones and populating the simulated light cones with high-redshift galaxies. \citet{Kennedy2024mlrec} then showed that this ML-recovery of ionized regions can shrink the volume needed to look galaxies, potentially increasing the efficiency of high-redshift surveys. On the cosmological side, \citet{2025PhRvD.111l3501T} stacked the ML-recovered ionized bubbles to search for deviations from isotropy, enabling an Alcock-Paczyński test with the SKA to measure the product of the angular diameter distance and the Hubble parameter to $\sim 2\%$ at reionization redshifts.

\subsection{Parameter Inference} \label{sec:infer}
One important avenue where ML is transforming data-driven research in cosmology and astrophysics is through ML-based regression and inference~\citep{Alsing:2019, Cranmer:2020, SpurioMancini2022, Villaescusa-Navarro:2022, Piras2024}. This links theory through model parameters directly to observations such as 21cm maps, i.e.~without the need of an explicit likelihood. However, we often rely on pre-defined observables and low-dimensional summaries, limiting the use of rich observational data. The observed 21cm signal exhibits non-Gaussianity, non-analytic likelihoods, and complex structure, and is made even more complicated by systematics and foregrounds. We therefore profit from network-based solutions for optimal and unbiased inference that go beyond oversimplified and biased summary statistics such as power spectra. Moving towards field-level approaches has necessitated the development and adaptation of newer inference techniques such as amortized inference.

Besides constraining the astrophysics of galaxies and the intergalactic medium~\citep{Park:2018ljd}, large-scale surveys such as mappings of the 21cm line with the SKAO enable studies of fundamental physics and cosmology~\citep{Liu:2020}. Fast simulation frameworks that are suitable for forward-modelling of synthetic observations and network training exist, for example, for generating 21cm 3D lightcones during cosmic dawn and reionization under different astrophysical~\citep{Murray2020} and cosmological~\citep{Heneka:2018} scenarios, complemented by smaller radiative hydrodynamical simulation datasets~\citep{Meriot:2023}. A significant modeling error remains, as well as a lack of end-to-end forward modeling pipelines that include all noise systematics, requiring inference methods able to transfer well and provide robust posterior estimation. In addition, a framework to bridge the transfer gap between simulations and observations, as well as between different simulators is crucial. 

For the 21cm signal, first inference applications of ML focused on regression of EoR parameters with dense neural networks (DNNs) from summaries such as 21cm power spectra~\citep{Shimabukuro_2017}, later extended to extraction of summaries such as the global 21cm signal in the presence of foregrounds, ionospheric effects, instrumental responses, and thermal noise~\citep{choudhury2020extracting, Choudhury_2021, Tripathi2024, Tripathi2024_Samp}, followed by regression from power spectra \citep{Choudhury2025} incorporating foreground effects~\citep{Choudhury_2022}. Since then, for parameter regression from full 21cm maps and light cones, convolutional neural networks (CNNs) have proven to work very well to summarise 21cm data for DNN regression of astrophysics and EoR properties~\citep{Gillet:2019,Hassan:2019, LaPlante:2019, Kwon:2020, Hassan_2020, Mangena:2020, Neutsch:2022,Prelogovic:2022,Hiegel_2023}. Besides aleatoric uncertainty (inherent noise in the data) captured by the variance in regression point estimates, epistemic uncertainty (model-related) in relation to the chosen network-model can in part be captured with Bayesian Neural Networks (BNNs). However, these methods do not provide full posteriors on the inferred parameters. When coupled with MCMC or variational inference though, BNNs were employed to approximate posteriors of the 21cm signal~\citep{Hortua_2020,Meriot:2025, Mahida_2025}. 

For full recovery and direct modelling of posteriors, neural density estimators (NDEs) are used within simulation-based inference (SBI) frameworks, both for direct inference from 21cm maps and light cones and for inference via fixed summary statistics such as 21cm power spectra. Here, NDEs are employed in various tasks, including neural likelihood estimation (NLE), neural posterior estimation (NPE) and neural ratio estimation (NRE); they offer powerful approaches for fast, robust inference from high-dimensional data and for high-dimensional posteriors, without having to rely on explicit likelihoods or overly simplistic summary statistics~\citep{Lueckmann:2019}. As flexible network-based density estimators these methods typically utilize coupling layers or spline-based transformations within normalizing flows~\citep{Rezende:2015} or autoregressive models like masked autoregressive flows~\citep{Papamakarios:2018}. For example, NPE directly estimates the posterior distribution $p(\mathbf{\theta}|x)$ from simulated data, and has seen wide use for SBI with synthetic SKAO data, based on realistic SKA-Low antennae configurations such as AA4 and the inclusion of thermal noise, systematics, and foregrounds.

\begin{figure}
    \centering
    \includegraphics[width=\columnwidth]{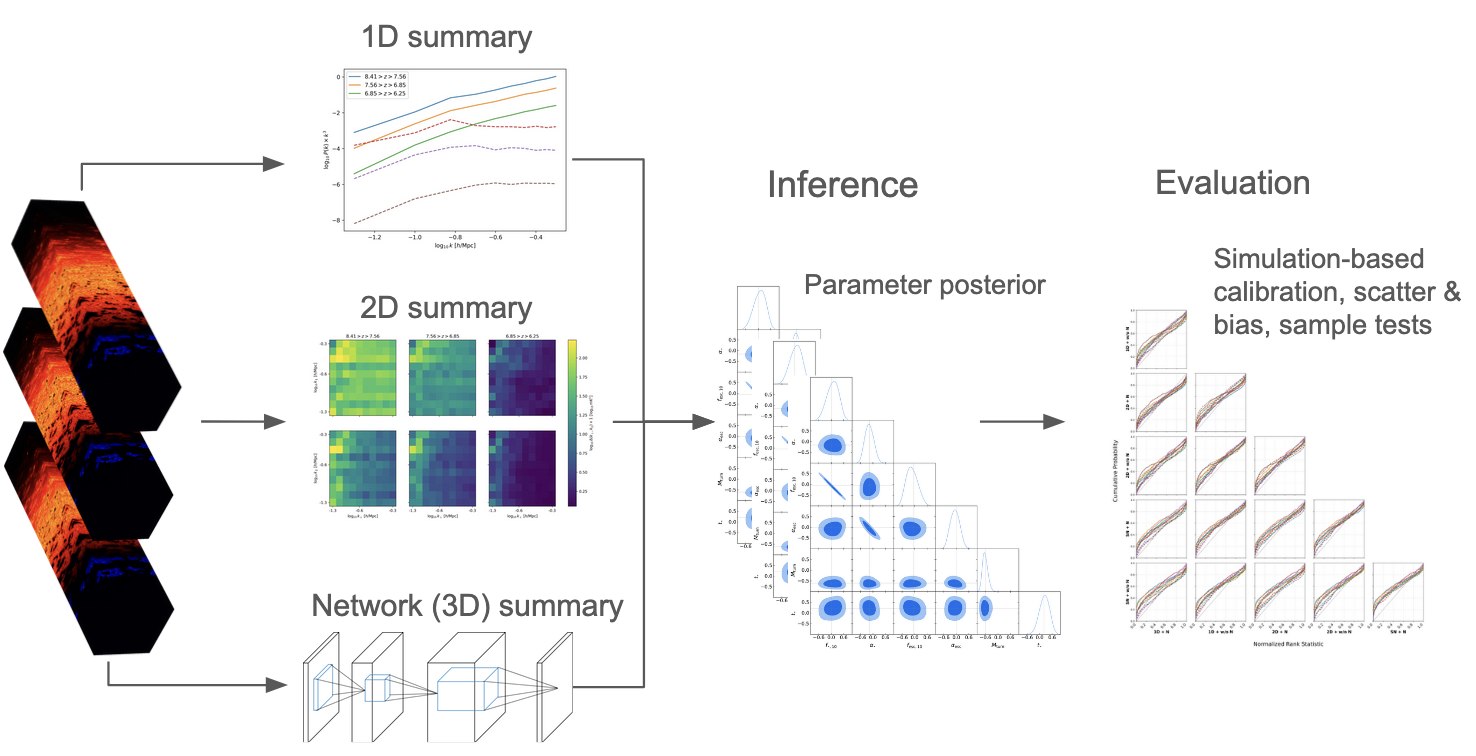}
    \caption{Possible workflows for likelihood-free inference (Sec.~\ref{sec:inference}): Once a database of 21 cm simulations is created, a suitable summary statistic (e.g. power spectra, higher-order statistics, or network-learned features) is extracted and used as input for simulation-based, likelihood-free inference based on neural density estimation. The resulting posteriors can be quantitatively evaluated using diagnostics such as simulation-based calibration and coverage tests.
    }
    \label{fig:4-1}
\end{figure}

Given the complexity and high dimensionality of the 21cm data expected from the SKAO, the choice of suitable summary statistics is crucial for effective SBI. As power spectra require less integration time to achieve sensitivity for signal detection, they are expected to precede high-fidelity imaging. Therefore, SBI for fixed summary statistics derived from synthetic SKA-Low data has been studied in form of NDE based on the 21cm one-dimensional spherically averaged power spectrum (1DPS) and the two-dimensional cylindrically averaged power spectrum (2DPS), e.g.~\citet{Saxena:2023,Greig:2024,Pietschke:2025}. 
As these studies have shown, SBI is able to derive EoR astrophysics and reconstruct the reionization history in an amortised and robust fashion. However, as for example~\citet{Zhao:2022b} and~\citet{Prelogovic:2023} have found, when faced with synthetic data that includes noise and systematics, non-Gaussian estimators are needed even in case of the 1DPS to avoid overconfident and biased posteriors. Furthermore, first results from the SKA Science Data Challenge 3b on inference (SDC3b\footnote{\url{https://sdc3.skao.int/challenges/inference}}), where the global ionization fraction at three different redshifts was to be blindly derived from 2DPS generated with two different simulators, point towards current posterior estimators to be susceptible to model misspecification both from signal and noise modelling; this has also been pointed out previously~\citep{Zhou:2022}. Furthermore, the work of \cite{CerardiGiri_2025} during the SDC3b shows that combining multiple summary statistics (2DPS, bispectrum or Betti numbers) can further enhance the extrapolation of global ionization fraction.

Finally, as an alternative to amortised inference with SBI based on a chosen summary statistic derived from simulated data, inference with for example MCMC becomes feasible when either leveraging ML for fast generative modelling of 21cm maps and light cones (see Sec.~\ref{sec:simulations}), or directly of statistics such as 21cm power spectra (see Sec.~\ref{sec:emulators}). 

However, 1D and 2D power spectra are by construction Gaussian statistics and thus limited in fully characterising the inherently non-Gaussian 21cm signal; higher-order statistics such as the bispectrum can add constraining power but typically suffer from low S/N requiring accurate noise modelling~\citep{Tiwari_2022, Mahida_2025}. Alternatively, SBI offers the possibility to directly infer high-dimensional posteriors from 21cm maps and light cones, circumventing the need for a predefined statistic and instead employ neural networks to learn, also non-Gaussian, summary representations directly from the data. For the global 21cm signal, this has been done without a summarisation step~\citep{Saxena:2024}. In the case of high-dimensional 21cm maps and light cones as expected for SKA-Low, usually an SBI framework with NDE has been combined with a summarisation network, or compressor to first reduce dimensionality and learn a summary statistic, for example based on convolutional or information-maximizing neural network architectures \citep{2018PhRvD..97h3004C}. This has been shown to provide a route to amortised, robust (unbiased) and optimal (neither over- nor underconfident) inference of astrophysical and cosmological information from synthetic SKA data~\citep{Prelogovic:2024, Schosser:2025, Schosser2025b}. In particular, data summarisation based on 3D convolutions has been shown to be efficient when dealing with 21cm light cones~\citep{Neutsch:2022,Zhao:2022a}. As has been noted in these studies, a joint optimisation of the network-learnt summary and posterior estimation is crucial in order to ensure robust and optimal parameter estimates. In addition, hybrid summaries and wavelets have emerged as an approach to combine physically motivated statistics with network-learned features to leverage both interpretability and data-driven optimization~\citep{Zhao:2023, Zhao:2024,Makinen:2024}. 

Figure~\ref{fig:4-1} depicts a workflow for ML-based inference of EoR model posteriors from SKA-Low data. We first derive either a fixed summary, such as a 1DPS or 2DPS, or a network-learnt summary as a lower-dimensional representation of high-dimensional SKA maps and light cones. As described in this section, within SBI frameworks for likelihood-free inference NDEs trained on simulated data directly estimate the high-dimensional parameter posterior. Also, fast emulators of simulations and summaries as described in Sec.~\ref{sec:simulations} and Sec.~\ref{sec:emulators} can be used for either likelihood-based Bayesian sampling or coupled to likelihood-free inference. Finally, accuracy and precision of derived posteriors can be quantitatively evaluated. For regression tasks, scatter plots that compare mean and variance for many realisations to determine bias are common. Quantitative approaches for inference include simulation-based calibration~\citep[SBC;][]{Talts:2018}, posterior coverage tests~\citep{Cook:2006}, and Tests of Accuracy with Random Points~\citep[TARP;][]{Lemos:2023}.
The majority of these methods are in principle also applicable to likelihood-based inference. Due to the intractable likelihood of the 21cm signal, for the SKA their use is enabled by either fast joint generation of parameters and signal data, or amortised posterior evaluation speed for SBI as compared to, for example, MCMC based on sampling 21cm simulations. One or more of the above-mentioned methods therefore are routinely used to judge posterior performance in 21cm inference with SBI. 

In a next step, studies of likelihood-free 21cm inference have recently explored routes to improved network-learnt summaries that are more robust than standard supervised approaches. The main motivation is an inherent problem for 21cm inference that arises due to differences observed across different 21cm forward models (model misspecification). Even when employing identical physical assumptions, distinct simulators can produce notably different outputs, introducing potential biases for an inference model trained on samples from only one simulator~\citep{Sooknunan:2024}. Related approaches such as Evidence Networks~\citep{2024MLS&T...5a5008J} demonstrate that neural estimators can perform amortized, simulation-based model comparison, offering a way to test and mitigate biases arising from mismatched simulators; and \citet{2025JCAP...08..004D} shows that neural density estimators such as continuous-time flow models can diagnose and quantify biases between forward models. Beyond discrepancies between simulators, inference methods must also handle variations in observational conditions, including instrumental noise models, resolutions, and systematics, factors crucially tested, for example, by SKA SDC3b. 
 Options so far explored for 21cm summaries are information-maximising networks~\citep{Prelogovic:2024}; whether they are maximally informative given the data is evaluated by comparing posterior variance against Fisher information or the Cramér–Rao bound. 
 Besides augmentation of training data, semi-supervised and self-supervised learning techniques based on masking and contrastive learning strategies, e.g.,~\citet{lecun2022path}, also provide a promising route toward addressing these robustness concerns. Indeed, self-supervised vision transformers trained on low-resolution 21cm light cones have been shown to generalize well to unseen higher-resolution, noised data and new model parameters~\citep{Ore:2025}, on the way to potentially exploit industry-sized models~\citep{Heneka:2025}.

Finally, it is very likely that synergies with other observables will ultimately be required to obtain the tightest constraints on our observables and to learn the most about CD/EoR. State-of-the-art inferences therefore combine with other probes such as UV luminosity functions that allow for the inclusion of JWST and HST observations, Thomson scattering optical depth to the CMB that allows for the inclusion of Planck observations, as well as full multi-probe analysis for example with CMB power spectra. A range of studies has already used ML for multi-probe joint inference with the SKA, precursor instruments such as HERA, and 21cm global signal measurements, using successfully NDE to evaluate joint likelihoods~\citep{Bevins:2023,Bevins:2024,Sims:2025}, or for joint NPE with conditional flow matching~\citep{Schosser2025b}; these studies have shown that SBI can robustly combine other probes with the 21cm signal in a likelihood-free way, leading to improved constraints on CD/EoR physics.

\section{Some Developing Trends}
\label{sec:autodiff}

In this chapter, we have discussed a number of advances in the application of ML to CD/EoR science. Given the infancy of this flavour of research, it would be fair to say that the role of ML in the SKA remains speculative for now, and may evolve considerably in the next few years.

However, it would be reasonable to guess that one low-level technique is likely to play a role: automatic differentiation (AD). AD is an emerging approach to solving difficult, high dimensional inverse problems, where one can compute the gradient of an objective function automatically with respect to its model parameters.
These gradients can then be used for optimization or sampling, and are particularly useful when working in high-dimensional parameter spaces.
Methods for computing gradients of a data model include: 1) numerical finite-difference approach, which is simple but contains numerical noise and can be prohibitively expensive in high dimensions; 2) symbolic regression, which computes exact gradients but is confined to working with closed-form analytic expressions, and 3) AD, which uses the chain rule of differential calculus to compute exact gradients of any series of numerical operations.
With access to gradients, one can employ a wide range of gradient-aware optimization techniques, such as 1st-order gradient descent algorithms and their derivatives (e.g. SGD, Adam), and 2nd-order gradient descent algorithms and their derivatives (e.g. BFGS, L-BFGS).
Furthermore, sampling the objective function, as is done with Markov Chain Monte Carlo approaches, can be made more efficient with gradients via algorithms like Hamiltonian Monte Carlo and its derivatives (e.g. NUTS).

If GPU-compute and large-scale datasets are the fuel of the AI revolution, then AD can be thought of as the train powering that revolution.
Indeed, it was the advent of flexible, user-friendly AD libraries that enabled the widespread adoption of deep learning \citep[][and references therein]{Gunes2018}.
While AD powers the training of neural networks, it can also be used to accelerate and scale-up our ability to solve traditional inverse problems, without the use of neural networks.
This general paradigm of programming with AD-enabled libraries is known as \emph{differentiable programming}, which has seen widespread adoption within the physical sciences.
The benefit here is the ability to leverage the inductive bias of our physics-based forward models, while still reaping the benefit of gradient-aware optimization and easy GPU portability afforded by high-level AD libraries (e.g. PyTorch, Tensorflow, JAX, Julia).

In 21\,cm cosmology, this has only recently begun to take hold.
\citet{Yattawatta2019} presented an approach for interferometric calibration using an L-BFGS optimizer coupled with an AD-enabled calibration forward model, demonstrating improved performance over standard calibration techniques.
\citet{EwallWice2022} presented a framework for the joint modeling of foregrounds and antenna beams in interferometric visibilities based on the principle of minimizing spectral structure.
This allows for the subtraction of beam-weighted foregrounds with minimal a priori assumptions about the structure of the beam.
Their forward model was wrapped in an AD-enabled library, and thus allowed for straightforward gradient-descent optimization and GPU-acceleration.
\citet{Cox:2023neq} took this a step further, showing how to leverage spectral redundancy in such an approach and thus improve the quality of antenna gain calibration. \citet{2025ApJS..278...25D} introduced a differentiable simulation for synchrotron from Galactic fields, which can be used for extract information from observation or integrated into AD-enabled pipelines.

\begin{figure}
    \centering
    \includegraphics[width=\columnwidth]{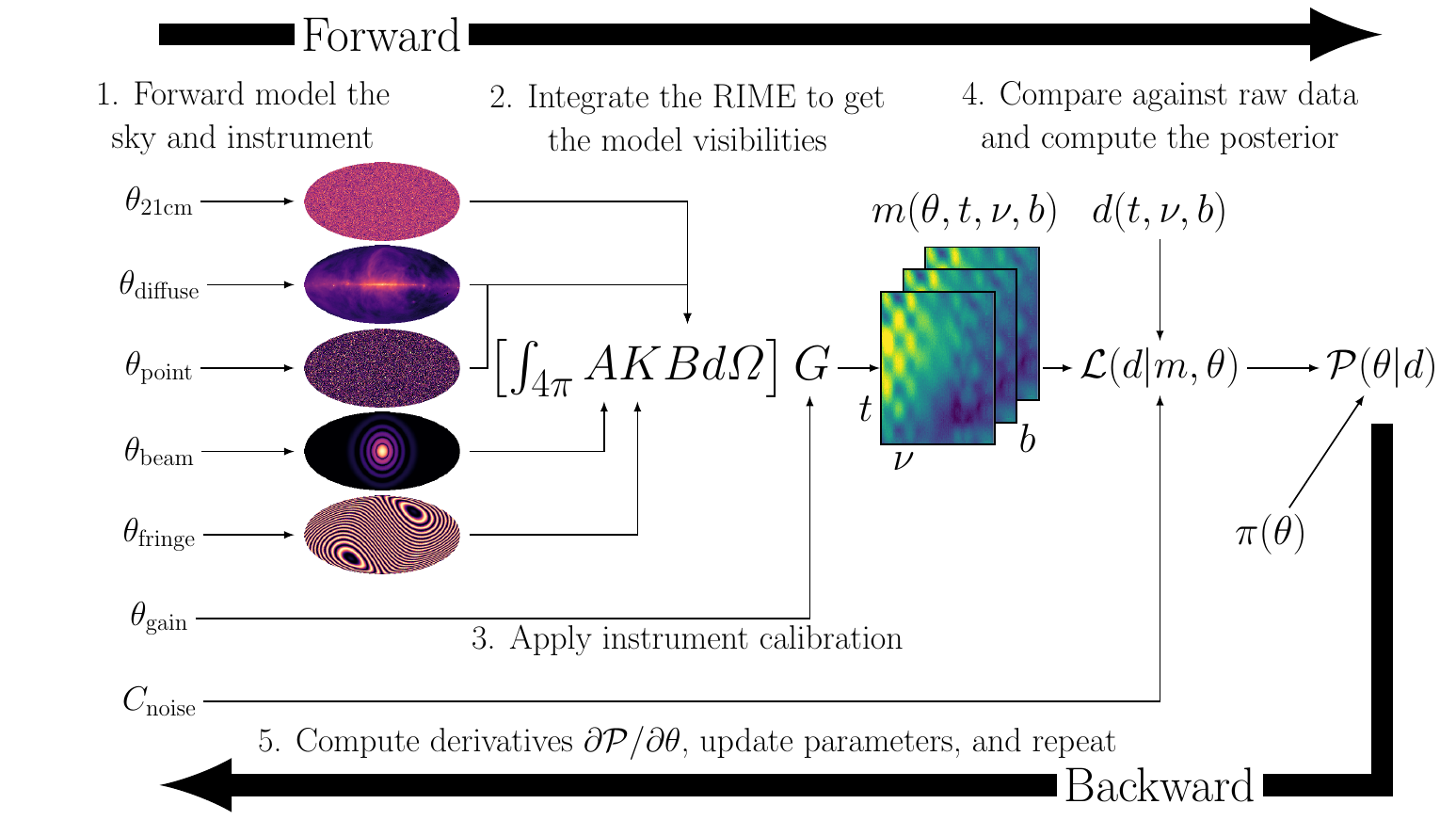}
    \caption{A differentiable, Bayesian forward model for 21\,cm cosmology from \citet{Kern2025}, highlighting the \emph{forward} operation of visibility simulation, and the \emph{backward} operation of reverse-mode backpropagation. Optimization and sampling of the (un-normalized) posterior distribution proceeds iteratively, using the computed gradients to move within the joint parameter space of the model. 
    }
    \label{fig:bayeslim}
\end{figure}

Recently, \citet{Kern2025} proposed an \emph{end-to-end} differentiable forward model for 21\,cm cosmological analysis.
In their forward model, signals are parameterized in their native space (e.g. on the sky), and then forward modeled to the interferometric visibilities via the measurement equation \citep{Smirnov2011}.
Wrapping this with an AD-enabled library allows one to differentiate backwards through the measurement process, and derive gradients of a visibility-based loss function with respect to sky pixels.
Coupled with an L-BFGS solver allows for joint optimization of foreground, instrument, and 21\,cm signal parameters.
Additionally, coupled with a Hamiltonian Monte Carlo algorithm \citep{Radford2011} allows for joint posterior density estimation and nuisance parameter marginalization.
This forward model is shown in \autoref{fig:bayeslim}.
Differentiable programming is crucial to making this feasible on large-scale datasets.
\citep{Kern2025} present a proof-of-concept for a scaled-down version of the HERA telescope, showing robust joint posterior density estimation between diffuse and point source foregrounds, a complex frequency and angle dependent antenna beam model, and the 21\,cm sky signal, with a total walltime of 16 hours spread across 4 GPUs.
Going forward, AD-enabled pipelines will help a variety of tasks within the standard 21\,cm workflow, such as antenna calibration, beam modeling, and foreground subtraction. 

Another important future prospect that cannot be ignored is the proliferation of modern AI tools based on foundation models, most notably large language models (LLMs). LLMs have achieved remarkable success in coding, summarizing papers, and synthesizing related literature. These tools are already being integrated into the daily workflows of many researchers, including many of those within the SKA 21-cm community. Looking ahead, foundation models will likely shape the paradigm of our research in two distinct ways: explicitly and implicitly.

Explicitly, foundation models can be trained to directly perform scientific tasks, including but not limited to map generation, parameter estimation, and source identification. There are two primary approaches to deploying such models. The first is training from scratch. There have been successful cases in astrophysics, particularly in fields where the number of catalogued astronomical objects reaches the hundreds of millions \citep[e.g.][]{AION-1,SpecCLIP,AstroCLIP}, enabling powerful performance in tasks like redshift and stellar parameter estimation. However, aggregating comparable volumes of training data remains highly challenging for 21 cm line intensity mapping and large-scale structure studies. Instead of training from scratch, the second approach is to fine-tune an existing foundation model. Fine-tuning allows one to retrain a general task foundation model on small dataset with limited training cost but achieve stronger performances on specific tasks (see e.g. \citet{finetuning-review} for a review). Because foundation models possess generalized capabilities learned across a wide range of domains, fine-tuning them with a relatively small amount of domain-specific data is often sufficient to achieve high performance on specialized tasks. For example, the work of \citet{Heneka:2025} fine-tuned existing foundation models, such as \texttt{Qwen}, on 21cmFAST simulations to analyze and generate 21-cm tomographic maps from astrophysical and cosmological parameters, confirming the viability of repurposing general-purpose foundation models for specialized cosmological applications. A true SKA-foundation model that can transfer between, e.g. training domains for different SKA-related tasks, similar to what has been done in Sec.~\ref{sec:inference}, remains to be built and tested, representing an interesting challenge for researchers to pursue.

Beyond these explicit applications, foundation models will also contribute to scientific output implicitly. By leveraging these tools, researchers can more effectively communicate scientific progress in papers and presentations, while AI-generated literature summaries will make complex research more accessible. Furthermore, the community will greatly benefit from AI coding assistants, which promise to streamline software development and data analysis pipelines. Recently, there are even trends of AI agents \citep[e.g.][]{agent}, constructing an automatic workflow of all above procedures with possible human interaction, accelerating the overall pace of scientific discovery.

\section{Summary and Future Prospects}
\label{sec:future}

In this chapter, we have examined applications of machine learning across the full analysis pipeline of CD/EoR science—from the observational data analysis to its theoretical interpretation. The breadth and depth of these applications suggest that ML will be a serious contender in the toolkit for realising the full scientific potential of the SKA in the long term. 

The methods described in Section~\ref{sec:obs} demonstrate a fundamental shift in how we approach observational data analysis. Traditional approaches to RFI mitigation, ionospheric correction, and gain calibration have relied on carefully engineered heuristics and statistical outlier detection. ML methods, on the other hand, can learn complex, non-linear patterns directly from data, adapting to the specific characteristics of each observation. For example, recent results from LOFAR demonstrate that improved gain calibration coupled with ML-based foreground mitigation can significantly improve data quality. As SKA-Low begins operations, we anticipate that similar methods will be essential and might transition from research prototypes as they currently are to operational components of standard analysis pipelines. The key challenge with such changes will be ensuring robustness and interpretability: ML models must not introduce biases or spurious features that could be mistaken for cosmological signal. Techniques such as explainable AI, uncertainty quantification, and rigorous validation against simulated datasets will be crucial in the coming years to build trust in these methods.

These advances in observational pipelines are complemented by equally significant progress on the theoretical side. The computational demands of EoR simulations represent a fundamental bottleneck to interpreting observations with forward modelling within an inference context. As discussed in Section~\ref{sec:theory}, field-level and summary statistic emulators offer a path forward, enabling inference at scales that would otherwise be computationally prohibitive. The next generation of emulators could incorporate several key advances: multi-fidelity approaches that combine cheap low-resolution simulations with expensive high-resolution runs to emulate large simulation boxes at high resolution, and physics-informed architectures that embed known symmetries and conservation laws. As emulators become more accurate and widely adopted, we may see a convergence between simulation-based and analytic modeling approaches, with differentiable simulators enabling end-to-end gradient-based inference through the full forward model.

Bridging theory and observation, Section~\ref{sec:inference} highlighted the shift toward likelihood-free, simulation-based inference methods that can extract information directly at the field level. This represents a shift away from a set of pre-defined summary statistics toward more optimal, data-driven approaches. Field-level inference will likely become routine, with NNs trained to map directly from observed data cubes to posterior distributions over astrophysical parameters. Hierarchical inference frameworks will enable simultaneous characterization of astrophysical populations and instrumental systematics, avoiding the artificial separation between ``signal" and ``nuisance" parameters. Multi-probe inferences will combine 21-cm data with complementary observations (e.g., see Synergies and Inference chapters; \citealt{Chakraborty01.2026.SKA, Acharya01.2026.SKA}) using ML methods designed for such synergistic analyses. A critical challenge will be ensuring that these methods produce well-calibrated uncertainties, particularly in the presence of model misspecification (see Inference chapter for more discussion on this; \citealt{Acharya01.2026.SKA}).

The development of autodifferentiable frameworks, as outlined in Section~\ref{sec:autodiff}, provides the technical foundation for many of these advances. Automatic differentiation enables gradient-based optimization and sampling in high-dimensional spaces, and the coming years will likely see the development of fully differentiable end-to-end pipelines that connect raw visibilities directly to astrophysical parameters, with gradients flowing through instrumental models, foreground components, and cosmological forward models. This will enable joint optimization over all aspects of the analysis.

While this chapter discusses ML applications to the individual parts of the CD/EoR science pipeline, the greatest impact may come from combining these methods. For example, a single differentiable pipeline might simultaneously perform RFI flagging, foreground removal, and parameter inference, with each component using some of the ML methods described in the corresponding subsections. Such end-to-end approaches would lead to extracting the maximal amount of information from upcoming observations.

A growing trend in machine learning is the development of foundation models: large, general-purpose models trained on diverse datasets that can be fine-tuned for specific tasks with relatively little additional data. In astronomy, this paradigm is also slowly beginning to take hold, with models pre-trained on large simulation suites or archival observations being adapted to new instruments or science cases. For CD/EoR science, foundation models could enable transfer learning from related domains such as galaxy surveys to SKA observations, potentially offering new ways to combine 21-cm data with complementary information.

Despite the impressive progress made so far, significant challenges that cannot be ignored remain, especially in terms of explainability and robustness. ML methods are often treated as "black boxes," making them difficult to understand, which is problematic for science in general, but becomes especially problematic when those methods fail. Ensuring that ML-driven methods are robust against adversarial inputs, out-of-distribution data, and model misspecification will require significant ongoing methodological development. 


The SKA is poised to be one of the defining radio observatories of the next decade. With its high sensitivity and high data rates, it will almost certainly be a key instrument in taking advantage of what ML has to offer to astrophysics. The challenge for the community in the coming years until then will be to develop, validate, and deploy ML methods responsibly, ensuring that they enhance rather than obscure our understanding of the Universe's first billion years.

\section{Author Contributions}
A. Liu, C. Heneka, and M. Bianco coordinated the writing of this chapter, organised the contributions and edited the content. They also provided a final general overview throughout the chapter, improving readability and formatting. We list the author's contributions for each section below.

\paragraph{Sec.~\ref{sec:intro}} A. Liu wrote this section.

\paragraph{Sec.~\ref{sec:obs}} C. Sui wrote the Sec.~\ref{sec:rfi}. S. K. Pal, A. Tripathi and A. Datta wrote Sec.~\ref{sec:ioneff} on ionospheric effects and provided a first draft of Sec.~\ref{sec:gain} on gain calibration. S. K. Giri wrote the first paragraph of the imaging section, Sec.~\ref{sec:imaging}, while Shulei Ni wrote the remaining paragraphs. A. Acharya primarily wrote Sec.~\ref{sec:foregroundmitigation} with contributions from S. K. Giri and D. Piras.

\paragraph{Sec.~\ref{sec:theory}} M. Bianco primarily wrote Sec.~\ref{sec:simulations}. Y. Mahida and S. Majumdar implemented the paragraph about the \texttt{CosmoUiT}, while C. Heneka implemented the LLMs part, and N. Kern wrote key details on emulation challenges and historical applications. S. K. Giri, D. Breitman and D. Piras provided general contributions to the section. D. Breitman coordinated the writing of Sec.~\ref{sec:emulators}, while A. K. Shaw, A. Tripathi, Y. Mahida and S. Majumdar primarily contributed to the fourth and fifth paragraphs. K. Diao contributed to the introduction of the map-level emulators in Sec.~\ref{sec:simulations}, focusing on GAN, diffusion, and model fine-tuning. X. Zhao contributed to the editing of Sec.~\ref{sec:theory}.

\paragraph{Sec.~\ref{sec:inference}} A. Liu wrote Sec.~\ref{sec:aux}. Y. Pietschke and C. Heneka primarily wrote Sec.~\ref{sec:infer}. A. Tripathi and Y. Mahida wrote the segments on the global signal and the application of BNN.  X. Zhao contributed to the editing of Sec.~\ref{sec:inference}. H.Shimabukuro contributed to the third paragraph at Sec 4.2.

\paragraph{Sec.~\ref{sec:autodiff}} A. Liu, N. Kern, and K. Diao wrote this section.

\paragraph{Sec.~\ref{sec:future}} D. Breitman wrote this section, with minor edits by A. Liu.
\bibliographystyle{abbrvnat-maxbibnames4}
\bibliography{chapter} 

\end{document}